\documentclass[10pt,conference]{IEEEtran}

\IEEEoverridecommandlockouts
\usepackage{cite}
\usepackage{amsmath,amssymb,amsfonts}
\usepackage{algorithmic}
\usepackage{graphicx}
\usepackage{textcomp}
\usepackage{xcolor}
\def\BibTeX{{\rm B\kern-.05em{\sc i\kern-.025em b}\kern-.08em
    T\kern-.1667em\lower.7ex\hbox{E}\kern-.125emX}}

\usepackage{subcaption}
\usepackage{fancybox}
\usepackage{float}
\usepackage{mdframed}
\usepackage{spverbatim}

\usepackage{xcolor}

\usepackage{url}
\usepackage{hyperref}

\usepackage{graphicx}
\usepackage{array}			
\usepackage{multirow}
\usepackage{tabularx}
\usepackage{booktabs} 

\usepackage{enumitem}

\usepackage{ifthen}
\newboolean{showcomments}
\setboolean{showcomments}{true} 
\ifthenelse{\boolean{showcomments}}
{\newcommand{\nb}[2]{
    \fcolorbox{gray}{yellow}{\bfseries\sffamily\scriptsize#1}
    {$\blacktriangleright$#2$\blacktriangleleft$}
  }
  
}
{\newcommand{\nb}[2]{}
  
}

\usepackage{xspace}

\newcommand{\comrat}{CoMRAT\xspace}
    
\begin{document}

\title{ \comrat: \\ Commit Message Rationale Analysis Tool 
\thanks{This research work is funded by the Fonds de Recherche du Qu\'{e}bec (B2X).}
}

\author{\IEEEauthorblockN{Mouna Dhaouadi}
\IEEEauthorblockA{
\textit{DIRO, Universit{\'e} de Montr{\'e}al}\\
Montr{\'e}al, Canada \\
0000-0001-9336-7714}
\and
\IEEEauthorblockN{Bentley James Oakes}
\IEEEauthorblockA{
\textit{GIGL, Polytechnique Montr{\'e}al}\\
Montr{\'e}al, Canada  \\
0000-0001-7558-1434}
\and
\IEEEauthorblockN{Michalis Famelis}
\IEEEauthorblockA{
\textit{DIRO, Universit{\'e} de Montr{\'e}al}\\
Montr{\'e}al, Canada  \\
0000-0003-3545-0274}

}

\maketitle

\begin{abstract}
In collaborative open-source development, the rationale for code changes is often captured in commit messages, making them a rich source of valuable information. However, research on rationale in commit messages remains limited.
In this paper, we present \comrat, a tool for analyzing decision and rationale sentences rationale in commit messages. \comrat enables a)
researchers to produce metrics and analyses on rationale information in any Github module, and 
b) developers to check the amount of rationale in their commit messages.
 A preliminary evaluation suggests the tool’s usefulness  and usability in both these research and development contexts.

 

\end{abstract}



\section{Introduction}
Reporting the \textit{rationale} behind a proposed code change is a necessary practice in collaborative open-source projects. In modern software development, where developers  rely on version control systems such as Git, rationale  information is often documented in the commit messages~\cite{dhaouadi2022end}. 
%
Although researchers have previously attempted to develop an understanding of developers’ rationale in open source software by studying chat messages~\cite{alkadhi2017react} or email archives~\cite{sharmaExtractingRationaleOpen2021}, research about rationale characteristics in the code commit messages of open-source projects is rather limited~\cite{tian2022makes, li2023commit, dhaouadi2024rationale}. 
Tian et al.~\cite{tian2022makes} and Li et al.~\cite{li2023commit} reported analyses regarding the quality of commit messages in terms of \textit{what} and \textit{why} information based on only five open source projects,  while our earlier work  examined developers' rationale specifically for the Out-Of-Memory Killer (OOM-Killer) module of the Linux Kernel~\cite{dhaouadi2024rationale}.

We present \comrat, a \textbf{Co}mmit \textbf{M}essage \textbf{R}ationale \textbf{A}nalysis \textbf{T}ool for studying developer’s rationale in the commit messages of any open source Github project. \comrat is a web application that includes: a) a \textit{Module Analyzer} for characterizing developer rationale in terms of its presence, impact factors, evolution and structure, and b) a \textit{Commit Message Analyzer} for promoting rationale-providing commits. 
A preliminary evaluation suggests \comrat's \textit{usefulness} (RQs.1,3) and \textit{usability} (RQs.2,4) for researchers and developers.

\section{Background}
\label{sec:background}

\subsection{The OOM-Killer dataset}

The Linux kernel is an extensive open-source project that has been developed collaboratively since 1991. 
Since 2005, Linux code patches have been organized as Git commits. Linux developers are encouraged to explain the motivation behind the commit and its impact on the kernel~\cite{love2010linux}. This community practice makes the kernel commits a valuable source of rationale information~\cite{dhaouadi2022end}.
%
The Linux project has several sub-projects that focus on different areas of the kernel, e.g.,  the `mm' folder contains code that focuses on memory management, the `fs' folder code for filesystems, etc. Each sub-project contains several modules. For instance, the `mm' sub-project contains the modules `oom\_kill', `slob' and  `migrate'. As these modules  have different concerns and committers, we treat each one as their own individual project.

In our previous work, we created an annotated, high-quality rationale dataset for the
OOM-Killer module~\cite{dhaouadi2024rationale}. 
We categorized sentences as \textit{Decision} (an action or a change that has been made), \textit{Rationale} (the reason for a decision or value judgment), and \textit{Supporting Facts} (narration of facts used to support a decision), and then quantitatively analyzed the resulting dataset to characterize
rationale in this subsystem. 
The next section details the analyses from~\cite{dhaouadi2024rationale} as implemented in \comrat. Section~\ref{sec:meth}  illustrates them with an  example.
\subsection{Rationale Analyses}
\label{sec:analyses}

The first set of analyses report on the presence of rationale in commit messages. We consider that a commit contains rationale if at least one of its sentences is labelled as Rationale. Specifically, we answer 
\textit{``How many commits contain rationale?''}
by introducing the \textit{rationale percentage} metric
and \textit{``How much of the commit contains rationale?''} by introducing the  \textit{rationale density} metric, as follows:\\

\begin{math}
\textit{rationale percentage} = \frac{\text{number of commits that contain rationale}} {\text{total number of commits}} \\ 
\end{math}


\begin{math}
\textit{rationale density} = \frac{ \text{number of sentences labelled as \textit{Rationale}} }{\text{total number of sentences in a commit} }\\
\end{math}

%

We also define the  $average \; rationale \; density$ for a specific module as follows:\\

\begin{math}
\textit{average rationale density} = \frac{ \sum_{\text{commits}} \text{rationale \; density} }{\text{number of commits that contain rationale} }\\
\end{math}

The second set of analyses concern the possible  factors impacting rationale, specifically the potential dependencies between the size of the commit and the developers experience, and the rationale density. To answer \textit{``Does the quantity of rationale reported depend on the commit message size?''}, we visualize the rationale density values versus the commit message size (i.e, number of sentences in a commit). 
To answer \textit{``Does the quantity of rationale reported depend on the developer experience?''}, we visualize the average rationale density per author (i.e, we compute the mean of the \textit{rationale density} of the commits of each author), along with the number of commits per author, as we consider the number of commits authored an indication of the developer's experience.

For rationale evolution, we answer \textit{``How does rationale evolve over time?''} by visualizing   the evolution of the average \textit{rationale density} and the average \textit{decision density} (calculated based on the decision-containing sentences) per year.

The final set of analyses concerns the structure of a commit message. That is, what sentence category order developers prefer when elaborating their commit messages. To answer \textit{``In what order do the categories mostly appear?''}, we visualize the distribution of the identified categories over the normalized positions of the sentences of the commit messages.


\subsection{Rationale Extractors}
In other work, 
we trained two binary BiLSTM classifiers on the OOM-Killer dataset (one for \textit{Decision}-containing sentences and one for \textit{Rationale}-containing sentences), and applied them on two different Linux modules: the \textit{slob.c\footnote{\url{https://api.github.com/repos/torvalds/linux/commits?path=mm/slob.c}}}
module of the \textit{mm} component and the 
\textit{button.c\footnote{\url{https://api.github.com/repos/torvalds/linux/commits?path=drivers/acpi/button.c}}} module of the \textit{drivers/acpi} subproject, and five other open source projects. We then validated their generalizability to these new contexts.

\section{\comrat Design, Implementation and usage}
\label{sec:meth}

We combine our previous implementation of the rationale analyses of the OMM-Killer module 
with our Bi-LSTM binary classifiers 
to make the analyses available for any Github project.
Having validated the generalizability of the extractors, 
we can apply them to other open source modules. 
In \comrat, we implement two analyzers that leverage the classifiers: a \textit{Module Analyzer}, and a \textit{Commit Message Analyzer}. The \textit{Module Analyzer} allows the user to apply the analyses defined above for a specific module. The \textit{Commit Message Analyzer} enables the user to enter a commit message and evaluate its quality. Our tool is packaged as a web application with two pages.
The only prerequisites to its  usage are a Github Username and a Github API token.

\subsection{\comrat Module Analyzer}
\label{sec:module_analyzer}

For the \textit{Module Analyzer}, we implement the following workflow: 1) The user enters the Github API URL of a specific module, then clicks the \textit{Start Module Analysis} button. 2) The analyzer downloads the commit messages through API calls.
3) The extracted messages are pre-processed similarly to the the training dataset of the classifiers.
4) The tool applies the classifiers on the pre-processed commit messages. 

\begin{figure}[htbp] 
\begin{minipage}{0.95\linewidth}
\begin{mdframed}
\textbf{\texttt{The mm/slob.c module}}

\begin{footnotesize}
\begin{verbatim}
Resulting dataset:

 % preview of the labelled dataset 

Number of commits: 146
Number of sentences: 833 
\end{verbatim}
\end{footnotesize}

\begin{small}
\begin{verbatim}
Distribution
\end{verbatim}
\end{small}

\begin{footnotesize}
\begin{verbatim}
Decision only sentences: 233
Rationale only sentences: 162
Decision & Rationale sentences: 252
No Decision and No Rationale sentences: 186
\end{verbatim}
\end{footnotesize}

\begin{small}
\begin{verbatim}
Word Clouds
\end{verbatim}
\end{small}

\begin{figure}[H]
 \begin{subfigure}{0.49\textwidth}
     \includegraphics[width=\textwidth]{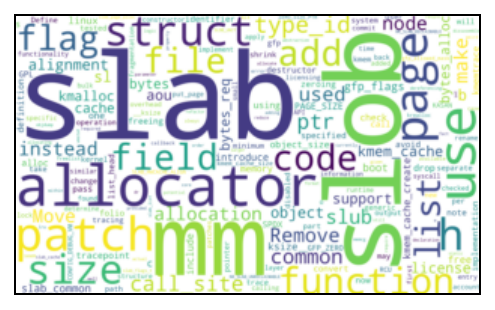}
 \end{subfigure}
 \hfill
 \begin{subfigure}{0.49\textwidth}
     \includegraphics[width=\textwidth]{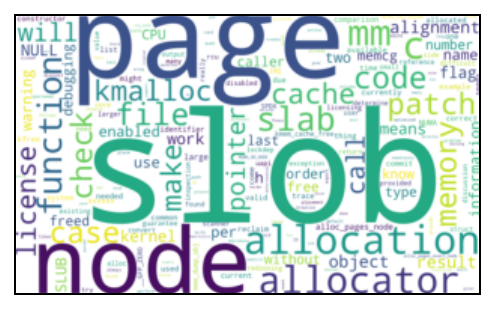}
 \end{subfigure}
\end{figure}

\begin{small}
\begin{verbatim}
Rationale Presence
\end{verbatim}
\end{small}

\begin{footnotesize}
\begin{verbatim}
Total Number of commits: 146
Number of commits that contain rationale: 124
Rationale Percentage: 84.93%
Average Rationale Density: 0.56
\end{verbatim}
\end{footnotesize}

\begin{small}
\begin{verbatim}
Rationale Factors
\end{verbatim}
\end{small}
\begin{figure}[H]
 \begin{subfigure}{0.49\textwidth}
     \includegraphics[width=\textwidth]{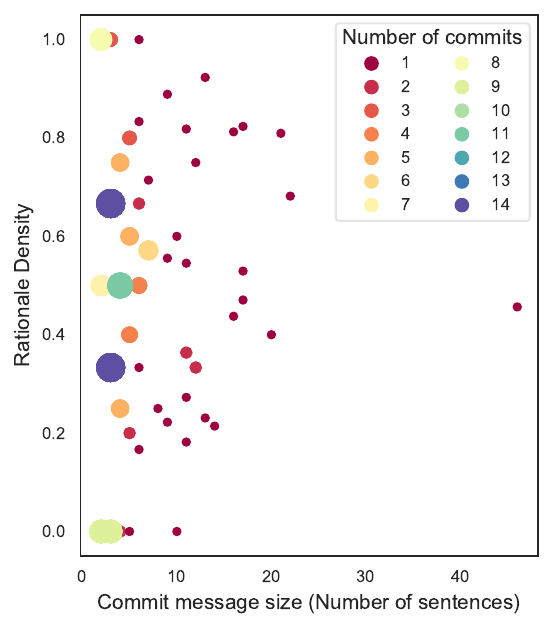}
 \end{subfigure}
 \hfill
 \begin{subfigure}{0.49\textwidth}
     \includegraphics[width=\textwidth]{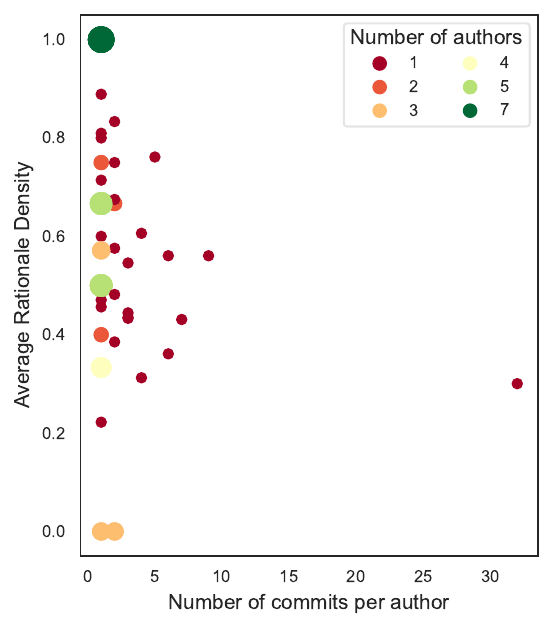}
 \end{subfigure}
\end{figure}

\begin{small}
\begin{verbatim}
Commit Message Structure
\end{verbatim}
\end{small}
\begin{figure}[H]
 \centering
     \includegraphics[width=0.75\textwidth]{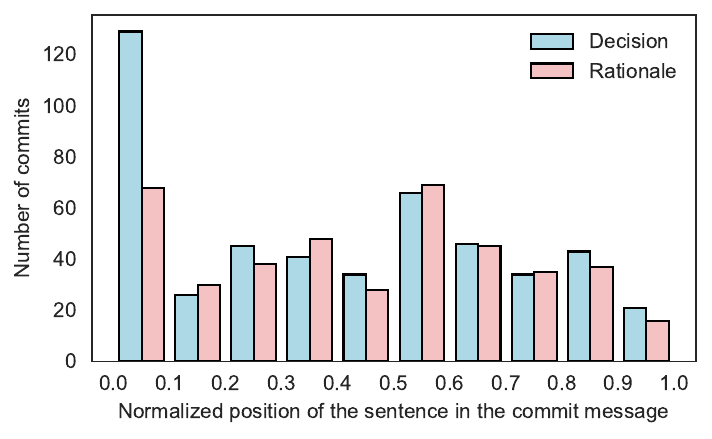}
\end{figure}
\begin{small}
\begin{verbatim}
Rationale Evolution
\end{verbatim}
\end{small}
\begin{figure}[H]
  \centering
     \includegraphics[width=0.95\textwidth]{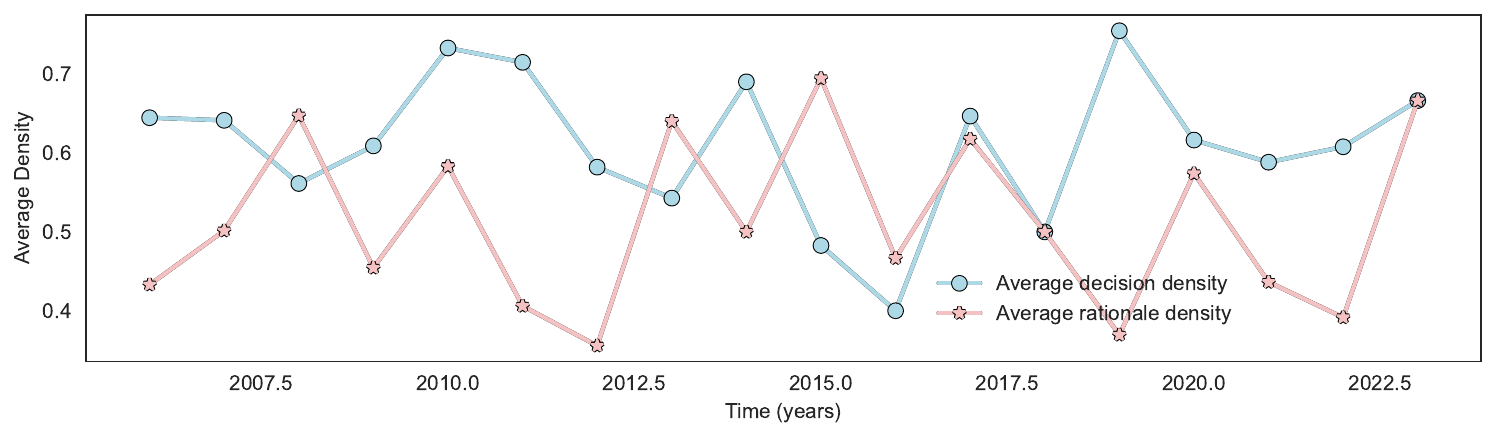}
\end{figure}
\end{mdframed}
\end{minipage}

\caption{\comrat analyses for the \textit{slob.c} module}
  \label{fig:slob_example}

\end{figure}

The output is a set of sentences labelled as Decision and/or Rationale. The tool also shows information about the distribution of the categories, and generates wordclouds from the most prominent words in each of the categories (without considering multi-labelled sentences). Note that when running the code locally, users can augment the built-in list of stop words of the wordcloud library\footnote{\url{https://amueller.github.io/word_cloud/}} with specific module-related keywords for better representations.
Finally, the tool analyzes the labelled dataset: it computes the metrics and  generates the figures. 
Optionally, the user can download the labelled dataset as a CSV file, or the generated figures as PDF files. Fig. \ref{fig:slob_example} shows the output relating to the \textit{slob.c} module.

\subsection{\comrat Commit Message Analyzer} 
\label{sec:commit_analyzer}
Here, the user inputs a commit message and clicks the \textit{``Start Commit Analysis''} button. \comrat pre-processes the message,   applies the classifiers, and outputs its sentences, labelled as Decision and/or Rationale. It also reports the \textit{number\_of\_sentences}, the \textit{rationale\_density} and the \textit{decision\_density} of the commit message. If the \textit{rationale\_density} is below the threshold of $0.5$, a warning message appears. If higher, a success message appears. Note that we are conducting further research to better determine this threshold value.







\subsection{Limitations}

\comrat is limited by GitHub’s API hourly rate limit (5000 authenticated requests per hour). This can be problematic 
for projects with longer commit histories, but can be overcome
by upgrading to GitHub Enterprise.
A second limitation is the non-optimal pre-processing we use~\cite{dhaouadi2024rationale}. 
In fact, due to improperly formatted commit messages, it is likely that the preprocessing may have been inaccurate or ineffective (e.g., not skipping blocks of raw source code from the message).
A third limitation is that we only consider commit messages, and not additional sources for rationale, such as issues~\cite{zhao2024drminer} or bug reports~\cite{lester2019identifying, rogers2015using}. This would require more downloaded data and further execution time.

\subsection{Installation and local execution}

\comrat is built with Python 3.11 and Streamlit\footnote{\url{https://streamlit.io/}} and only needs
specific libraries installable through \texttt{pip} in order to work. 
The source code is publicly available~\cite{replication_package}. 
To install, users can download the source and, navigating to the \textit{Rationale-Analyses-Tool} folder, create a virtual environment to install the required libraries and run the tool locally:
\begin{verbatim}
    pip install -r requirements.txt
    streamlit run tool.py
\end{verbatim}
    

\subsection{Potential Uses}

The \textit{Module Analyzer} could be used to create datasets of commits with rationale from open source Github projects, unearthing valuable rationale knowledge that could be reused in future development projects. In industrial settings, this would enhance decision-making and productivity. Structural analysis  
may 
reveal structural patterns and identify common commit messages structures that could be used as guidelines or encouraged via automated tools. 
In development contexts, this would ensure well-structured commit messages, or suggest improvements before merging commits. For example, the \textit{Commit Message Analyzer} allows the user to evaluate the quality of their commit message, e.g., by scoring their commit information against the module to which they are contributing to. This ensures that a desired level of rationale information is obtained, which is particularly valuable in open source projects for onboarding newcomers  and retaining design information.

The \textit{Module Analyzer} could also be used by researchers for in-depth investigation of  the rationale in Github projects.
Subsequent studies could reuse the research questions in Section~\ref{sec:analyses} or include a comparison between modules in terms of rationale information. For instance, Table~\ref{tab:runtime} shows the \textit{rationale percentage} and the \textit{average rationale density} for five Linux modules, 
suggesting stable rationale distribution and generalizability of prior findings from the OMM-Killer module~\cite{dhaouadi2024rationale}.
The generated datasets could also support future research on how developers express their rationale, or regarding other factors influencing rationale in commit messages,  e.g., the number of reviewers  or the author's affiliation.
%
%

 \section{Preliminary Evaluation}



We conducted a preliminary evaluation of \comrat to assess usefulness (RQs.1,3) and usability (RQs.2,4). Complete materials / anonymized data are in a replication package~\cite{replication_package}.

\subsection{Module Analyzer}

\paragraph{Usefulness (RQ.1)} 
To get an expert evaluation, we asked a Software Engineering researcher with more than 10 years of experience to test the \textit{Module Analyzer}. 
Then, we asked him to fill a form that contains a Likert scale about the usefulness and the usability of the tool overall, and six rating questions (with a scale of 1 to 5 stars) about the usefulness of each of the analyses. 
%
%
The researcher \textit{Strongly Agreed} that the labelled dataset and the generated analyses are useful for researchers, with a minimum of four stars of five for the six analyses. He rated the Word Clouds analysis with 2/5 stars as he found the images were informative but could not be reused, and suggested frequency tables instead. 
He also described the kinds of research that  might benefit from \comrat:
\begingroup
\addtolength\leftmargini{-0.1in}
\begin{quotation}
``Many types of MSR papers could benefit from such tools. I can think of using commit messages (with decisions and rationales) to enhance code review and bug localization. 
[..]
Automatic bug fixes (or at least learning from past bug fixes) would also likely benefit.''
\end{quotation}
\endgroup

\paragraph{Usability (RQ.2)} The expert researcher \textit{Agreed} that the tool is easy to use.
We also report in the first three  columns in Table~\ref{tab:runtime} the execution time required to get the data through API calls, load and apply the classification models, execute the analyses, and generate the figures. The time was measured on a laptop with Intel Core 1.30 GHz and 16 GB of memory. 
The results suggest that running \comrat locally is  feasible.





\begin{table}[h]
    \centering
    \begin{tabular}{|c|p{1.3cm}|p{1.3cm}||p{1.3cm}|p{1.6cm}|}
    \toprule
       Module  & Number of Commits  & Execution Time & Rationale Percentage (\%)& Average \newline Rationale Density (\%) \\
       \midrule
       mm & 404 &  2m 24s & 98.9   &  61.4  \\  
       slob  &  146  &  ~~~~~34s  & 84.9 & 56.0 \\ 
       button  &  110   & ~~~~~26s  & 90.9 & 61.9 \\ 
       FS  &  15    & ~~~~~13s & 86.7 & 46.5\\ 
      migrate  &  666 & 8m~~0s  & 94.3  & 63.0  \\ 

       \bottomrule
    \end{tabular}
    \caption{Information about different Linux modules}
    \label{tab:runtime}
\end{table}

\subsection{Commit Message Analyzer}

We conducted a preliminary user study to evaluate the usefulness and usability of the \textit{Commit Message Analyzer}. 

\textit{Study design}. 
Our study includes a  labelling task and a post-study questionnaire. The first author defined the Decision and Rationale categories and introduced the sentence-based labelling task using three commit examples. Then, the participants were asked to manually label the sentences of six other commit messages, three with rationale density higher than 0.5 and three lower. 
%
The labelling task was carried out to prompt participants to critically analyze and reflect on the messages. 
The  first author  then introduced the \textit{Commit Message Analyzer} and asked the participants to apply it on the six commit messages. Participants then filled the questionnaire.

Our questionnaire comprises 29 English-language questions. It
includes two multi-choice demographics questions about participants’ background (current position and experience with Git development),
and four Likert scale questions per commit about the quantity and helpfulness of  rationale information in the commit and about the tool's ability to identify that information and provide helpful feedback messages. 
Finally, we include three Likert scale questions about the helpfulness of the labelling produced by the \textit{Commit Message Analyzer} as well as its usability and impact overall.


\textit{Participants}.
To date, we have conducted the study with five participants
with appropriate command of English to understand the commit messages.
We use convenience sampling as we recruit participants from the university we are affiliated with. 
%
Our participants are all graduate students (40\% Master’s and 60\% PhD) with a minimum of two years with Git development. 
They frequently write commit messages as part of their academic activities and collaborative research projects.

\textit{Initial Study Results.} 
 
\textit{a) Usefulness (RQ.3):}
All users (100\%) either \textit{Agreed} or \textit{Strongly Agreed} that the tool provides helpful labelling.
For five out of six commit messages, at least 60\% either \textit{Agreed} or \textit{Strongly Agreed} that the feedback message was helpful.

\textit{b) Usability (RQ.4):}
All users (100\%) either \textit{Agreed} or \textit{Strongly Agreed} that \textit{the tool was easy to use},
and that \textit{it encourages adding rationale to commits}.


\vspace{0.2cm}
\setlength{\fboxsep}{7pt}%
\ovalbox{%
\begin{minipage}{7.5cm}

\textbf{Evaluation Summary.}
From this initial study, we claim that \comrat is both \textit{useful (RQs.1,3)} and \textit{usable (RQs.2,4)} for researchers and developers. 

\end{minipage}}

\section{Related Work}

Automatically mining developers rationale has
recently attracted research interest.
Alkadhi et al. experimented with machine learning models to extract  rationale elements
(decision, issue, alternative, pro-argument, con-argument) from developers' chat messages~\cite{alkadhi2017react}. 
 Rogers et al.~\cite{rogers2015using} used 
 linguistic features while Lester et al.~\cite{lester2019identifying} experimented with evolutionary  algorithms to optimize the
feature sets to improve rationale extraction from Chrome bug reports.
Sharma et al. tried to extract rationale from Python Enhancement Proposals using heuristics~\cite{sharmaExtractingRationaleOpen2021}. 
Kleebaum et al. proposed automatic rationale classification from Jira issues and commit messages using machine learning~\cite{kleebaum2021continuous}.
Zhao et al.   mined design rationales from developers discussions in open-source issue logs
using Large Language Models and heuristics. They also investigated the usefulness of the extracted information for automated program repair~\cite{zhao2024drminer}.
These prior works differ from us as  
they only focus on the extraction and do not  analyze developer's rationale. Also, none of this prior research has examined the rationale in open-source commit messages.  

Researchers have also studied commit message quality\cite{tian2022makes, li2023commit}. 
Tian et al.  define what constitutes a ``good'' commit message~\cite{tian2022makes} by studying five open-source projects. They found that it should summarize \textit{what} was changed, and
describe \textit{why} those changes are needed,  and proposed a good-message identification tool. 
Li et al.~\cite{li2023commit}
continued this research by considering link contents in addition to the commit messages while training
classifiers for the automatic identification of good commit messages.
They also studied the commit quality evolution over time, and the correlation between defect proneness and the quality of the commit message.

Similar to us, these
researchers considered rationale information in the  commit messages of open source projects and proposed analyses that include the temporal evolution aspect and the factors that influence rationale.
Different from us, 
their  analyses only consider the evolution of the existence of \textit{what} (decision) and \textit{why} (rationale) information over time, while we study the evolution of their quantities. They also only focus on the correlation with defect proneness while we study factors that might influence rationale. Also, they do not consider the structure (order) in which the \textit{what} and \textit{why} information appear. 
Finally, previous work does not provide their classifiers as a tool applicable on any Github module as we do.

 \section{Conclusion}
We present \comrat, a tool that a) provides researchers with labelled datasets and insights about rationale information in  any Github project and b) assists developers in writing rationale-aware commits. 
Preliminary evaluation 
indicates \comrat's usefulness and usability for researchers and developers.
In the future, we plan to package the \textit{Commit Message Analyzer} as a Github Bot to be integrated directly into the development process, by scoring the rationale information in pull request commits.



\bibliographystyle{IEEEtran}
\bibliography{rationale.bib, linux}


\end{document}